\documentclass[twocolumn,final,aps,prb,showpacs,superscriptaddress,amsmath,amssymb,amsfonts,floatfix]{revtex4-2}
\usepackage{epsfig}
\usepackage[dvipsnames,usenames]{color}
\usepackage{xcolor}
\usepackage{tcolorbox}
\usepackage{amsmath}
\usepackage{comment}
\usepackage{array}
\usepackage{multirow}
\usepackage{tabularx}
\usepackage{float}
\usepackage[colorlinks=true,bookmarks=true, pdfborder={0 0 0},linkcolor=blue,urlcolor=blue,     breaklinks=true,bookmarksnumbered=true]{hyperref}
\usepackage[title]{appendix}
\usepackage[normalem]{ulem}
\usepackage{subfigure}
\emergencystretch=\maxdimen
\begin{document}

\title{Perpendicular electric field induced competing $s^\pm$- and $d$-wave pairings in La$_3$Ni$_2$O$_7$ thin film}

\author{Yongping Wei}
\affiliation{Institute for Quantum Science, School of Physical Science and Technology, Soochow University, Suzhou 215006, P. R. China}

\author{Xun Liu}
\affiliation{Institute for Quantum Science, School of Physical Science and Technology, Soochow University, Suzhou 215006, P. R. China}

\author{Fan Yang}
\affiliation{School of Physics, Beijing Institute of Technology, Beijing 100081, P. R. China}

\author{Mi Jiang}
\affiliation{Institute for Quantum Science, School of Physical Science and Technology, Soochow University, Suzhou 215006, P. R. China}
\affiliation{State Key Laboratory of Surface Physics and Department of Physics, Fudan University, Shanghai 200433, P. R. China}

\begin{abstract}
Inspired by the possibility that superconducting properties may be altered by applying a perpendicular electric field in the Ruddlesden–Popper (RP) bilayer nickelate La$_3$Ni$_2$O$_7$ thin film, we investigated the imbalanced two-orbital bilayer Hubbard model with a layer potential bias using dynamical cluster quantum Monte Carlo calculations. Focusing on the pairing symmetry evolution with the potential bias between layers for the undoped, hole-doped, and electron-doped regimes, we found that the $s^\pm$-wave pairing, originating from the $d_{z^2}$ orbital at high temperature regime as well as $d_{x^2-y^2}$ orbital at lower temperatures, is suppressed by the potential bias; while a possible pairing symmetry transition from $s^\pm$-wave to $d$-wave pairing occurs, driven by the interlayer orbital mismatch and the transfer of electrons into the $d_{x^2-y^2}$ orbitals. Our large-scale many-body calculations align with the previous expectation from weak-coupling methods and provide further insight into the superconducting mechanism in RP nickelates.
\end{abstract}

\maketitle

\section{Introduction}
Since the discovery of high-temperature superconductivity (SC) with a critical temperature $T_c\sim$ 80 K in pressurized La$_3$Ni$_2$O$_7$~\cite{WM2023,YHQ2024,PRX2024,JMST2024,CJG2024}, Ruddlesden–Popper (RP) phase nickelates have attracted significant interest in the field of condensed matter physics. In particular, nickelates with higher critical temperatures, such as La$_2$SmNi$_2$O$_7$ ($T_c\sim90$ K)~\cite{90K} and Nd-doped La$_3$Ni$_2$O$_7$ ($T_c\sim100$ K)~\cite{100K}, have been successfully synthesized. Furthermore, experiments have not only revealed that another RP phase multilayer nickelate, La$_4$Ni$_3$O$_{10}$, exhibits superconductivity with $T_c\sim30$ K under pressure~\cite{43101,43102,43103}, but have also confirmed that thin-film La$_3$Ni$_2$O$_7$ on SrLaAlO$_4$ substrates at ambient pressure shows high-temperature SC with $T_c\sim40$ K~\cite{film1,film2}. Moreover, (La,Pr)$_3$Ni$_2$O$_7$ films on the same substrate have achieved $T_c\sim60$ K through a giant-oxidative atomic-layer-by-layer epitaxy method in an extreme nonequilibrium growth regime~\cite{60Kfilm}. In fact, $T_c$ can be enhanced by the in-plane compressive strain controlled via different substrates~\cite{substrate}. Additionally, an incomplete dome of $T_c$ as a function of hole-doping level has been observed upon doping with strontium (Sr)~\cite{Sr327}.

To understand the difference in SC between cuprates and nickelates, it is essential to uncover the mechanism of SC in RP nickelates, in addition to the infinite-layer nickelates~\cite{Ni1121,Ni1122} at ambient pressure. There are two main microscopic theories regarding the superconducting mechanism of La$_3$Ni$_2$O$_7$. One considers the $d_{z^2}$ orbital as the origin, where electrons pair via strong interlayer spin antiferromagnetic fluctuations, while phase coherence of these electron pairs is established through $d_{x^2-y^2}$–$d_{z^2}$ hybridization~\cite{YYF}. The other suggests that Hund's coupling between the two orbitals plays a key role in transferring magnetic correlations from interlayer $d_{z^2}$ electrons to interlayer $d_{x^2-y^2}$ electrons, such that $s^\pm$-wave pairing SC instability arises from the $d_{x^2-y^2}$ orbital~\cite{WCJ2024,YFarxiv}. In fact, it remains unclear whether the pairing symmetry is $s$-wave~\cite{Sandreev,CZY1,CZY2} or $d$-wave~\cite{Dandreev}. Previous studies have shown that a perpendicular electric field can induce rich phenomena in the RP nickelate. In the strongly correlated limit, the slave-boson mean-field (SBMF) approach~\cite{YFNC} revealed a pairing symmetry transition from $s^\pm$-wave to $d$-wave, accompanied by a higher $T_c$. Another RPA study of a two-band model~\cite{twoband} and the further simplified effective single-orbital model~\cite{YHZhang} predicted a pair density wave state. Furthermore, the perpendicular electric field can also induce Fermi arcs or nodal points and modulate the gap along the diagonal of the Fermi surface~\cite{TZhou}.

In this paper, we employ the dynamical cluster approximation on the two-orbital bilayer Hubbard model with an additional layer potential bias $E$ to mimic the perpendicular electric field, as illustrated in Fig.~\ref{lattice}, to examine the above idea. Our analysis reveals that the $s^\pm$-wave pairing, which originates from the $d_{z^2}$ orbital at high temperature regime and shifts to the $d_{x^2-y^2}$ orbital at lower temperatures, is suppressed by the potential bias. In contrast, the $d$-wave pairing on the $d_{x^2-y^2}$ orbital is enhanced and might exhibit a dome-shaped behavior as $E$ increases. These findings are consistent with a possible electric-field-driven pairing-symmetry transition. Furthermore, we observe that the $s^\pm$-wave pairing is generically strengthened by electron doping.

This paper is organized as follows: Sec.II introduces the model and methodology; Sec.III presents the competing $s^\pm$- and $d$-wave pairings in the undoped, hole-doped, and electron-doped cases; and Sec.IV provides a summary and outlook.

\section{Model and Method}
\subsection{Imbalanced bilayer two-orbital model}
\begin{figure}
\psfig{figure=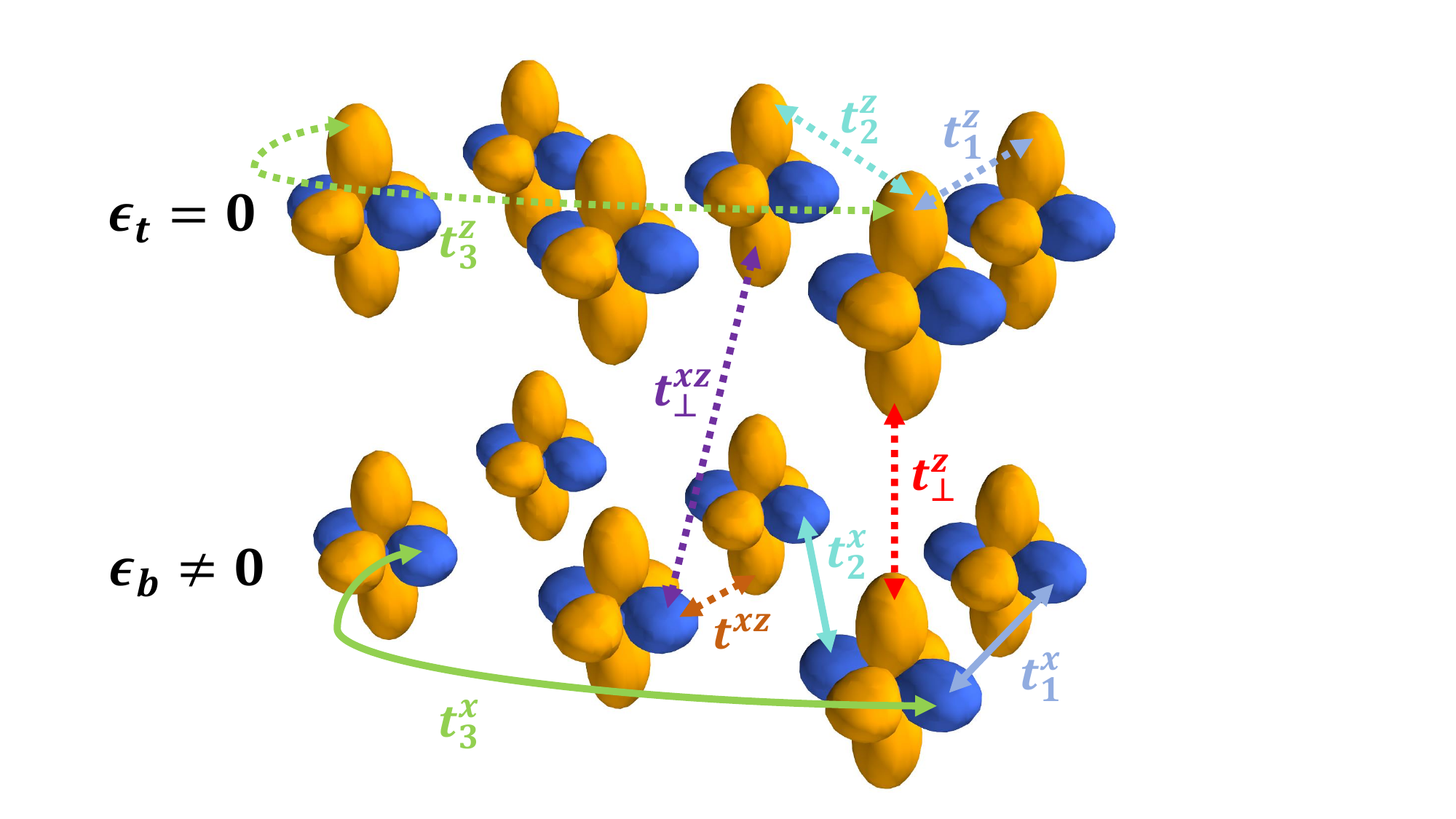, height=5.6cm,width =.53\textwidth, clip}
\caption{Schematic diagram of the imbalanced bilayer two-orbital model. The solid and dashed arrows represent the hoppings for the $d_{x^2-y^2}$ and $d_{z^2}$ orbitals, respectively. The $d_{x^2-y^2}$ interlayer hopping $t_\perp^{xx}$ is neglected to simplify the Hamiltonian. Brown and purple arrows indicate the intralayer and interlayer inter-orbital hoppings. An additional on-site energy on the bottom layer $\epsilon_b \neq 0$ is introduced to simulate a layer potential bias $E=\epsilon_b-\epsilon_t$, where $\epsilon_t=0$.}
\label{lattice}
\end{figure}

We consider the imbalanced bilayer two-orbital model for the thin-film La$_3$Ni$_2$O$_7$ on a two-dimensional square lattice, as shown in Fig.~\ref{lattice}, with the Hamiltonian

\begin{align}
    H& =  H_0+H_U \notag \\
    H_0 &= \sum_{ i j \sigma m l\nu} t_\nu^{x/z} c_{i ml \sigma}^\dagger c_{j ml \sigma}+t^{xz}\sum_{\langle ij \rangle \sigma mll'}c_{i ml \sigma}^\dagger c_{j ml' \sigma} \notag \\
    &\quad+ t_\perp^{xz} \sum_{\langle\langle ij \rangle\rangle ll'\sigma} c_{i 1l \sigma}^\dagger c_{j 2l' \sigma}+t_\perp^{zz} \sum_{i \sigma} c_{i 1z \sigma}^\dagger c_{i 2z\sigma}\notag \\
    &\quad + (\epsilon^{x/z} - \mu)\sum_{iml\sigma} n_{i ml \sigma}+\epsilon_b \sum_{iml\sigma} n_{i ml \sigma}\delta_{1m} + \mathrm{H}.c. \notag \\
    H_U &= U \sum_{i ml} n_{i ml \uparrow} n_{i ml \downarrow} + U' \sum_{imll'\sigma\sigma'} n_{i ml\sigma} n_{i ml' \sigma'} \notag \\
    &\quad+ (U'-J) \sum_{imll'\sigma} n_{i ml\sigma} n_{i ml' \sigma} \notag
\end{align}
where $c_{i \sigma}^{\dagger}$ ($c_{i \sigma}$) creates (annihilates) an electron with spin $\sigma$ (= $\uparrow$, $\downarrow$) in the $m$-th (= 1, 2) layer and orbital $l = (d_{x^2-y^2}(x), d_{z^2}(z))$ at site $i$. The terms $t_\nu^{x/z}$ include the nearest-neighbor, next-nearest-neighbor, and next-next-nearest-neighbor hoppings $t_1^{x/z}$, $t_2^{x/z}$, and $t_3^{x/z}$ for the $d_{x^2-y^2}$ and $d_{z^2}$ orbitals. The terms $t^{xz}$ and $t_\perp^{xz}$ denote the intralayer and interlayer inter-orbital hoppings, respectively. The tight-binding parameters we adopted are listed in Table~\ref{paras}, as reported in Ref.~\cite{YFNC,YDXpara}. We set the interlayer $t_\perp^{xx} = 0$ for simplicity and set the on-site Coulomb interaction $U = 4.0$ eV, the inter-orbital Coulomb interaction $U' = U - 2J = 2.4$ eV with the Hund's coupling $J = U/5 = 0.8$ eV as typical values. The chemical potential $\mu$ tunes the desired average density $n$. 

To simulate the effect of a realistic perpendicular electric field, we introduce an additional tunable site energy $\epsilon_b \neq 0$ on the bottom layer (see Fig.~\ref{lattice}). This gives rise to a layer potential bias $E = \epsilon_b - \epsilon_t$, where $\epsilon_t$ is set to zero.

\begin{table}[H]
\caption{The tight-binding parameters for the imbalanced bilayer two-orbital model of La$_3$Ni$_2$O$_7$ thin film, as shown in Fig.~\ref{lattice}. The additional on-site energy on the top layer is $\epsilon_t = 0$. All values are in units of eV.}
\centering
\begin{tabular}{cccc}
\hline\hline
\makebox[2.0cm]{$t_1^x$} & \makebox[2.0cm]{$t_2^x$} & \makebox[2.0cm]{$t_3^x$} & \makebox[2.0cm]{$t_\perp^{xx}$} \\ [0.7ex]
\hline
\makebox[2.0cm]{$-0.445$} & \makebox[2.0cm]{$0.06$} & \makebox[2.0cm]{$0.057$} & \makebox[2.0cm]{$0$} \\ [0.7ex]
\hline\hline
\makebox[2.0cm]{$t_1^z$} & \makebox[2.0cm]{$t_2^z$} & \makebox[2.0cm]{$t_3^z$} & \makebox[2.0cm]{$t_\perp^{zz}$} \\ [0.7ex]
\hline
\makebox[2.0cm]{$-0.131$} & \makebox[2.0cm]{$-0.015$} & \makebox[2.0cm]{$-0.011$} & \makebox[2.0cm]{$-0.503$} \\ [0.7ex]
\hline\hline
\makebox[2.0cm]{$t^{xz}$} & \makebox[2.0cm]{$t_\perp^{xz}$} & \makebox[2.0cm]{$\epsilon^x$} & \makebox[2.0cm]{$\epsilon^z$} \\ [0.7ex]
\hline
\makebox[2.0cm]{$0.221$} & \makebox[2.0cm]{$-0.031$} & \makebox[2.0cm]{$0.756$} & \makebox[2.0cm]{$0.389$} \\ [0.7ex]
\hline\hline
\end{tabular}
\label{paras}
\end{table}

\subsection{Dynamical cluster approximation (DCA)}
The dynamical cluster approximation (DCA) with the continuous-time auxiliary-field (CT-AUX) quantum Monte Carlo (QMC) cluster solver~\cite{Hettler98,Maier05,code,GullCTAUX} is employed to numerically solve the bilayer two-orbital model. As a well-established quantum many-body numerical method, DCA evaluates various physical observables in the thermodynamic limit by mapping the bulk lattice problem onto a finite cluster embedded in a mean-field bath in a self-consistent manner~\cite{Hettler98,Maier05}, which is achieved through the convergence between the cluster and coarse-grained single-particle Green's functions, where the coarse-graining is performed over a patch of the Brillouin zone around the cluster momentum $\mathbf{K}$.

In this manner, the short-range interaction within the cluster are treated exactly using various numerical techniques, such as CT-AUX in the present study; while longer-ranged physics is approximated by a mean field hybridized with the cluster. Therefore, increasing the cluster size systematically approaches the exact result in the thermodynamic limit. The finite cluster size effectively approximates the entire Brillouin zone by a discrete set of $\mathbf{K}$ points, such that the self-energy $\Sigma(\mathbf{K},i\omega_n)$ is constant within the patch around a particular $\mathbf{K}$ and takes the form of a step function across the full Brillouin zone.

In general, quantum embedding methods including DCA exhibit a milder sign problem compared to finite-size QMC simulations due to the presence of the mean field. However, limited by the complexity of the model, most of our calculations were performed using an $N_c = 8$ ($=4 \times 2$)-site DCA cluster that includes the $(\pi,0)$ and $(0,\pi)$ points in order to access lower temperatures and capture the $d$-wave pairing symmetry.

\subsection{Bethe-Salpeter equation (BSE)}

The superconducting properties can be studied by solving the Bethe–Salpeter equation (BSE) in its eigenvalue form in the particle-particle channel~\cite{Maier2006,scalapino2007numerical}
\begin{align} \label{BSE}
-\frac{T}{N_c}\sum_{K'}
\Gamma^{pp}(K,K')
\bar{\chi}_0^{pp}(K')\phi_\alpha(K') =\lambda_\alpha(T) \phi_\alpha(K),
\end{align}
where $\Gamma^{pp}(K,K')$ denotes the irreducible particle-particle vertex of the effective cluster problem, with $K = (\mathbf{K}, i\omega_n)$ combining the cluster momenta $\mathbf{K}$ and Matsubara frequencies $\omega_n = (2n+1)\pi T$, and $\phi_\alpha(K)$ represents the eigenvector obtained by solving the BSE for the pairing symmetry of type $\alpha$.

The coarse-grained bare particle-particle susceptibility
\begin{align}\label{eq:chipp}
	\bar{\chi}^{pp}_0(K) = \frac{N_c}{N}\sum_{k'}G(K+k')G(-K-k')
\end{align}
is obtained via the dressed single-particle Green's function,
\begin{align}
G(k)\equiv G({\bf k},i\omega_n) =
[i\omega_n+\mu-\varepsilon_{\bf k}-\Sigma({\bf K},i\omega_n)]^{-1}
\end{align}
where $\mathbf{k}$ belongs to the DCA patch surrounding the cluster momentum $\mathbf{K}$. The non-interacting dispersion is
\[
\varepsilon_\mathbf{k}
=
\renewcommand{\arraystretch}{1.5}
\begin{bmatrix}
E_\mathbf{1}^{x} & \quad E^{xz} & \quad t_\perp^{xx} & \quad E_\perp^{xz}\\
E^{xz} & \quad E_\mathbf{1}^{z} & \quad E_\perp^{xz} & \quad t_\perp^{zz}\\
t_\perp^{xx} & \quad E_\perp^{xz} & \quad E_\mathbf{2}^{x} & \quad E^{xz} \\
E_\perp^{xz} & \quad t_\perp^{zz} & \quad E^{xz} & \quad E_\mathbf{2}^{z}
\end{bmatrix}
\]
with

\begin{align}
E_{m}^{x/z}&=\epsilon^{x/z}+\epsilon_{b}\delta_{1m}+2t_1^{x/z}(\cos k_x+\cos k_y) \notag\\
&\quad+4t_2^{x/z}\cos k_x\cos k_y+2t_3^{x/z}(\cos 2k_x+\cos 2k_y) \notag \\
E^{xz}&=2t^{xz}(\cos k_x-\cos k_y) \notag \\ 
E_{\perp}^{xz}&=2t_\perp^{xz}(\cos k_x-\cos k_y) \notag
\end{align}

The chemical potential is $\mu$, and $\Sigma(\mathbf{K}, i\omega_n)$ is the cluster self-energy. In practice, we calculate 32 or more discrete points for both the positive and negative fermionic Matsubara frequency $\omega_n=(2n+1)\pi T$ mesh for measuring the two-particle Green's functions and irreducible vertices. Therefore, the BSE Eq.~\eqref{BSE} reduces to an eigenvalue problem of a matrix of size $(128N_c)\times (128N_c)$. 

The pairing operator is defined as
\begin{align}
\Delta_\alpha^{\dagger} &= \frac{1}{\sqrt{N}} \sum_{\mathbf{K}} g_\alpha(\mathbf{K}) c_{\mathbf{K}}^\dagger c_{-\mathbf{K}}^\dagger.
\end{align}
Here, $\alpha$ denotes the pairing symmetry and the $s^\pm$-wave and $d$-wave symmetries are represented by the form factors $g_{s^\pm}(\mathbf{K}) = \cos \mathbf{K_z}$ and $g_d(\mathbf{K}) = \cos \mathbf{K_x} - \cos \mathbf{K_y}$, respectively.

To systematically investigate the interplay between the additional layer potential bias and doping, we analyze the temperature evolution of the BSE eigenvalue $\lambda_\alpha(T)$ ($\alpha = s^\pm, d$). The observation of BCS-like logarithmic temperature dependence would indicate behavior analogous to that observed in the $d$-wave superconducting phase of the conventional single-band Hubbard model in the overdoped regime~\cite{Maier2019,eigenlog}. Conversely, the emergence of linear or exponential temperature dependence would suggest non-BCS pairing fluctuations, similar to those characteristic of the pseudogap regime in the single-band model~\cite{Maier2019}.
Unfortunately, the fermionic sign problem restricts the achievable temperature range, making it challenging to reliably extract the extrapolated $T_c$ from the temperature dependence of $\lambda_\alpha(T)$. Therefore, the relative magnitude of the eigenvalues serves as a more practical metric for assessing the pairing strength in our current study.

\begin{figure}
\psfig{figure=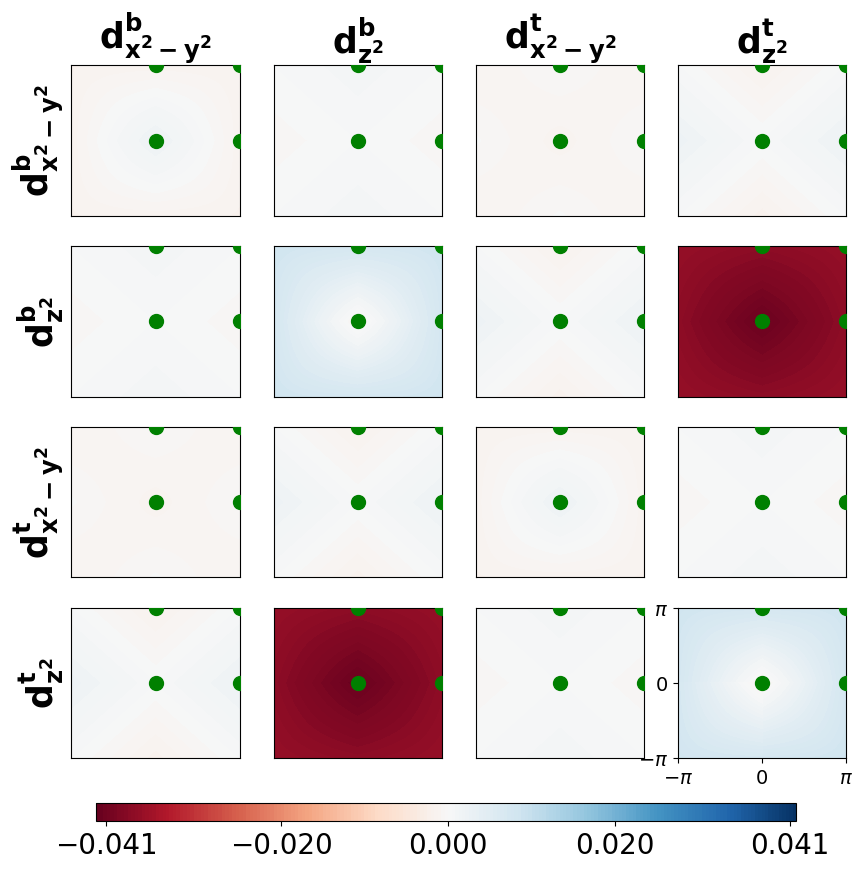,width=0.23\textwidth,clip}
\psfig{figure=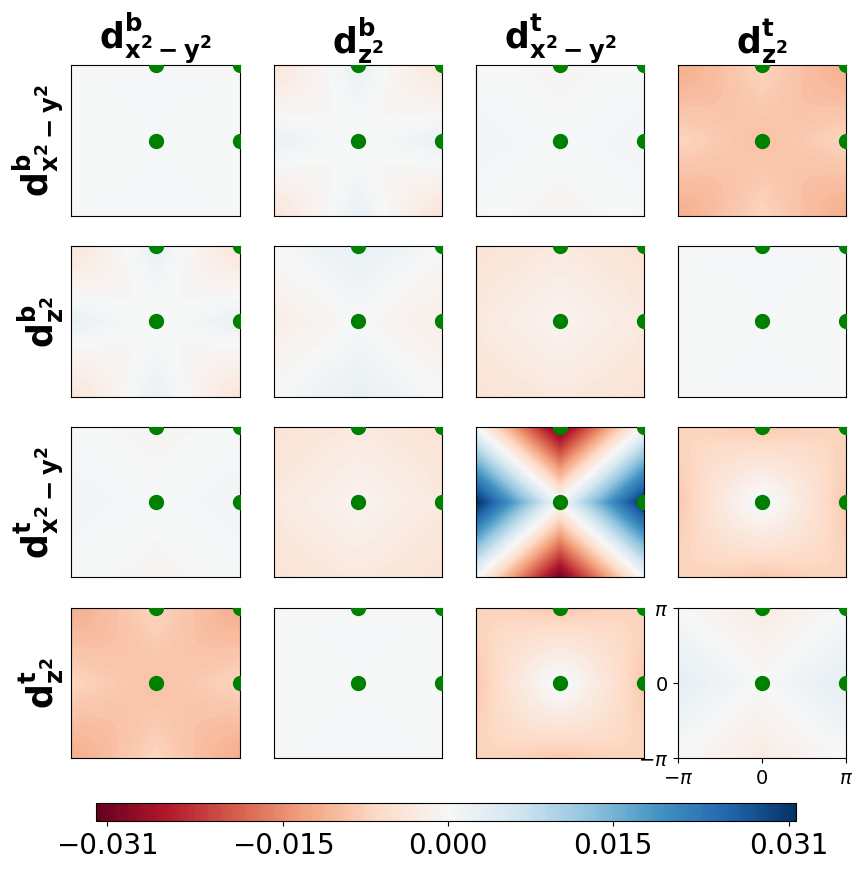,width=0.23\textwidth,clip}
\caption{The leading BSE eigenvector in orbital space at $E = 0.0$ eV (left) and $1.0$ eV (right) at $T = 0.1$ eV in the undoped $n=3.0$ case, which corresponds to $s^\pm$-wave and $d$-wave pairing symmetry, respectively.} 
\label{vectorn3}
\end{figure}

\begin{figure*}
\psfig{figure=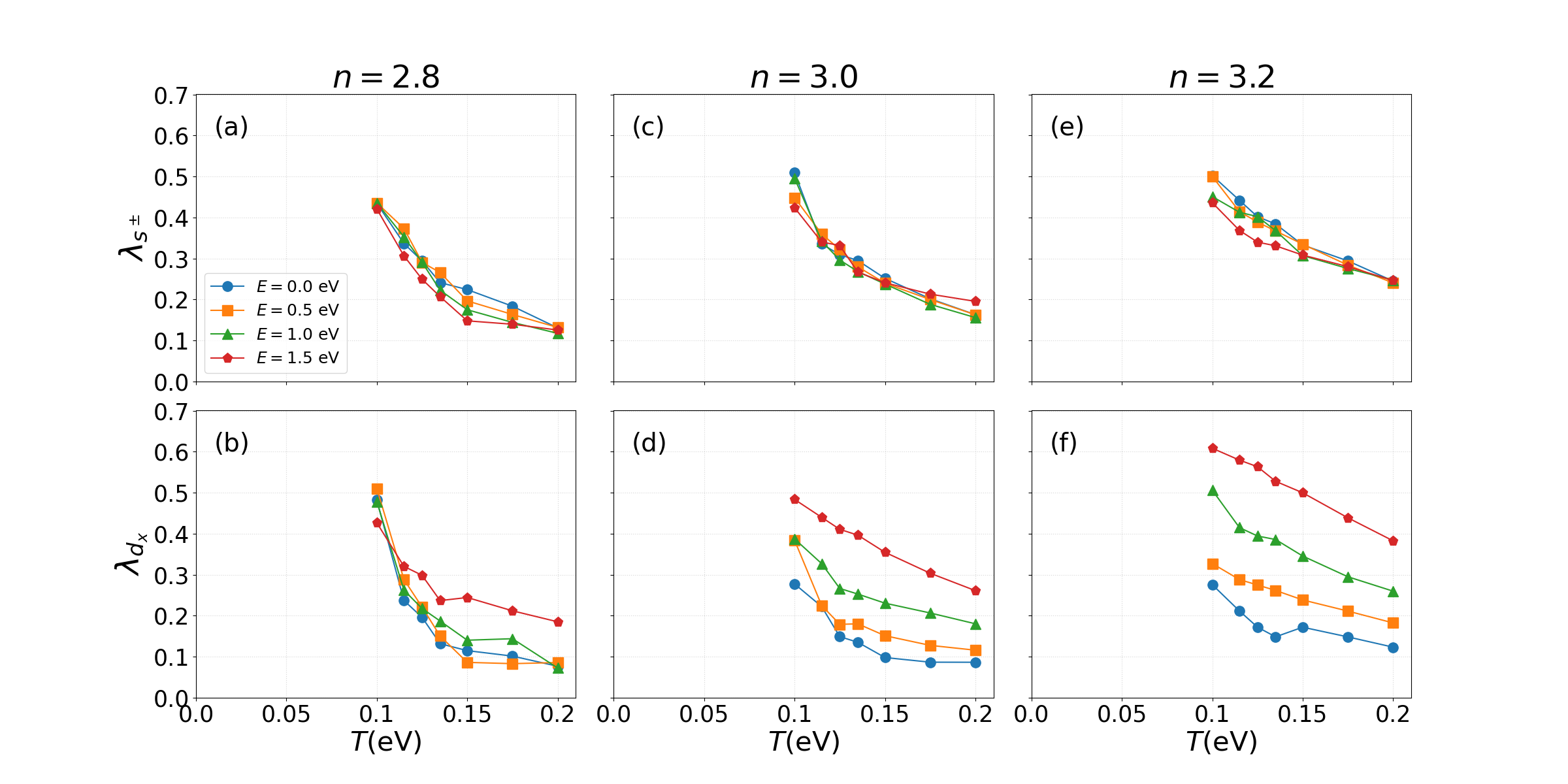,width =1.\textwidth, clip}
\caption{Temperature evolution of the BSE eigenvalues for $s^\pm$-wave (upper panels) and $d$-wave (lower panels) pairing under a series of layer potential bias $E = 0.0, 0.5, 1.0$, and $1.5$ eV for the three characteristic electron densities $n = 2.8$ (left), $3.0$ (middle), and $3.2$ (right).}
\label{value}
\end{figure*}

\subsection{Pair-field susceptibility}
We also decompose the pair-field susceptibility into two orbital-resolved components~\cite{WW2015,WW2025}: $P^x$ for $d_{x^2-y^2}$ orbital and $P^z$ for $d_{z^2}$ orbital, in order to examine their respective roles for the two different pairing symmetries.

Following the usual DCA formalism~\cite{Hettler98,Maier05,code,GullCTAUX}, the pair-field susceptibility is defined as

\begin{align}
P^{x/z}(T)&=\frac{T^2}{N^2_c}\sum_{K,K'} \notag \\ &\sum_{l_1l_2l_3l_4}g_{\alpha}(\mathbf{K})\delta_{l_1l_2}\bar{G}^{pp}_{l_1l_2l_3l_4}(K,K')g_{\alpha}(\mathbf{K'})\delta_{l_3l_4}
\end{align}
where $\bar{G}^{pp}_{l_1l_2l_3l_4}(K,K')$ is the coarse-grained four-point two-particle Green's function for the orbital components ($l_1l_2l_3l_4$), which can be evaluated by the BSE as

\begin{align}
\bar{G}^{pp}_{l_1l_2l_3l_4}&(K,K')= \notag \\
&\bar{G}^{pp,0}_{l_1l_2l_3l_4}(K,K')\delta_{K,K'}+\bar{G}^{pp,\Gamma}_{l_1l_2l_3l_4}(K,K')
\end{align}

Thus, we can define the non-interacting and interacting parts of the two components~\cite{maiernpj}
\begin{align}
P^{x/z}_0(T)&=\frac{T^2}{N^2_c}\sum_{K,K'}\delta_{K,K'} \notag \\
&\sum_{l_1l_2l_3l_4}g_{\alpha}(\mathbf{K})\delta_{
l_1,l_2}\bar{G}^{pp,0}_{l_1l_2l_3l_4}(K,K')g_{\alpha}(\mathbf{K'})\delta_{
l_3,l_4}\\
P^{x/z}_\Gamma(T)&=\frac{T^2}{N^2_c}\sum_{K,K'} \notag\\
&\sum_{l_1l_2l_3l_4}g_{\alpha}(\mathbf{K})\delta_{
l_1,l_2}\bar{G}^{pp,\Gamma}_{l_1l_2l_3l_4}(K,K')g_{\alpha}(\mathbf{K'})\delta_{
l_3,l_4}
\end{align}

In fact, the non-interacting part $P^{x/z}_0(T)$ is obtained by multiplying the coarse-grained bare particle-particle susceptibility in Eq.~(\ref{eq:chipp}) by the form factor, and then decomposing it into two orbital-resolved quantities. The more important interacting part $P^{x/z}_\Gamma(T)$ can be obtained as
\begin{align}
\bar{G}^{pp,\Gamma}_{l_1l_2l_3l_4}(K,K')&=\frac{T}{N_c}\sum_{K,K'}\sum_{l_5l_6l_7l_8}\bar{G}^{pp,0}_{l_1l_2l_5l_6}(K,K') \notag\\
&\times\Gamma_{l_5l_6l_7l_8}(K,K')\bar{G}^{pp,0}_{l_7l_8l_3l_4}(K,K')
\end{align}
where $\Gamma_{l_5l_6l_7l_8}(K,K')$ is the reducible four-point vertex~\cite{maier2001}. Notably, the absence of the momentum points $\mathbf{K}=(\pi,0)$ and $\mathbf{K}=(0,\pi)$ in the $N_c=1\times2$ cluster precludes an analysis of the $d$-wave pair-field susceptibility.


\begin{figure}
\psfig{figure=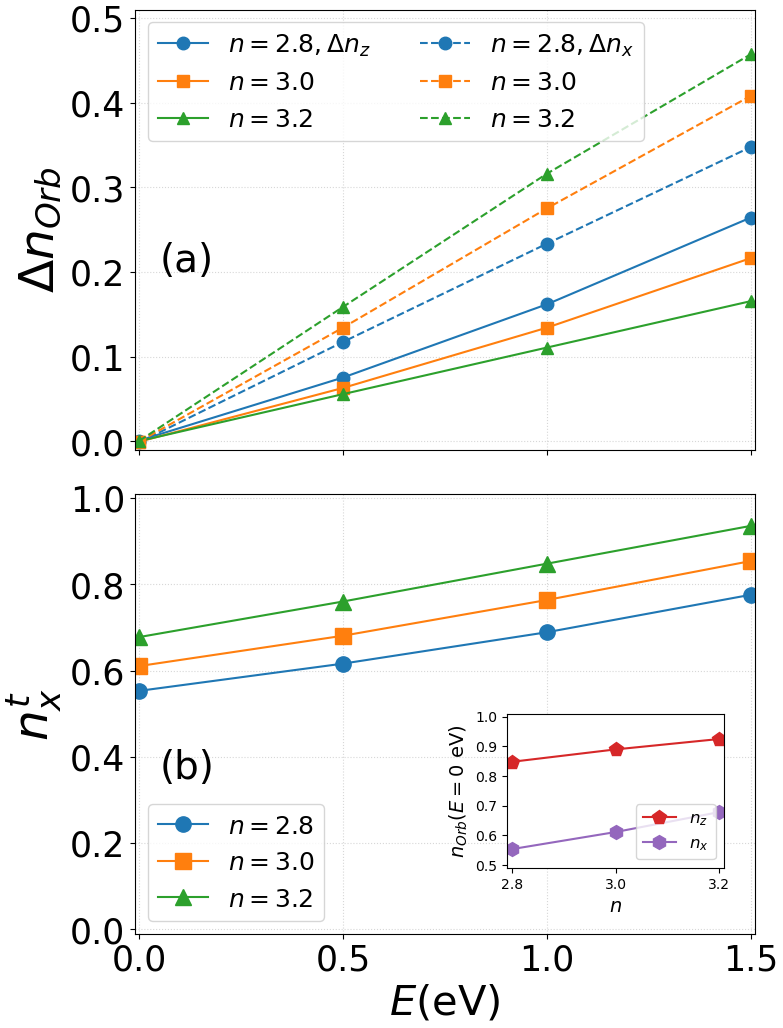,width=0.45\textwidth,clip}
\caption{(a) The density discrepancy $\Delta n_z = n^b_z - n^t_z$ and $\Delta n_x = n^b_x - n^t_x$ as a function of the potential bias $E$. (b) The density $n^t_x$ of the top $d_{x^2-y^2}$ orbital, which hosts the $d$-wave pairing, as a function of the potential bias $E$. The inset shows the evolution of the density distribution ($n_x = n^b_x = n^t_x$ and $n_z = n^b_z = n^t_z$) with total density $n$ at $E = 0.0$ eV, the gray dashed line indicates the optimal filling. The temperature is $T = 0.1$ eV.}
\label{filling}
\end{figure}

\section{Results}

For a systematic study of the two-orbital bilayer model in the presence of a layer potential bias, our results are divided into two parts. Part I is devoted to the competing $s^\pm$- and $d$-wave pairing symmetries via the BSE eigenvalues for the $N_c = 4 \times 2$ cluster. Part II investigates the temperature dependence of the orbital-resolved pair-field susceptibility for the same cluster. Furthermore, the $N_c = 1 \times 2$ cluster results are included to extend the analysis of $s^\pm$-wave pairing at much lower temperatures. 

\subsection{BSE eigenvectors and eigenvalues}
In the undoped case, the leading eigenvectors in the orbital basis at the characteristic $E = 0.0$ and $1.0$ eV at our lowest reliable $T = 0.1$ eV are compared in Fig.~\ref{vectorn3}. For $E = 0.0$ eV, the interlayer pairing on the $d_{z^2}$ orbital is apparently strongest, indicating the interlayer $s^{\pm}$ pairing symmetry, which is similar to that obtained in previous work using the FLEX method~\cite{pair}. 

At $E = 1.0$ eV, the upper and lower layers are inequivalent and consequently the interlayer pairing with $s^{\pm}$-wave symmetry largely weakens. Alternatively, some dominant features from $d_{x^2-y^2}$ orbitals emerge, which clearly displays the intralayer $d$-wave pairing in the top layer. It is worth noting that previous studies revealed the possibility that the $d_{x^2-y^2}$ orbital can also support $s^\pm$-wave pairing via the Hund's coupling~\cite{WCJ2024,YF2,YFarxiv,YFNC}. In fact, it is noteworthy that our recent investigation of Ni$_2$O$_9$ cluster adopting exact diagonalization also emphasized the important role of $d_{x^2-y^2}$ orbital as well as the in-plane Oxygen~\cite{ED1,ED3}. Nevertheless, no clear signal is observed in our simulations at the lowest temperature of $T=0.1$ eV with $N_c=4\times2$. Furthermore, the calculated pair-field susceptibility presented later, obtained at a lower temperature but with a smaller $N_c=1\times2$, suggests that the $d_{x^2-y^2}$ orbital plays a dominant role in the $s^{\pm}$-wave SC. 

Besides, Fig.~\ref{vectorn3} reveals that the top $d_{z^2}$ orbital also exhibits $d$-wave pairing, suggesting the role of its hybridization and/or Hund's coupling with $d_{x^2-y^2}$ orbital, which even propagates to the bottom layer via strong interlayer hybridization between $d_{z^2}$ orbitals. In fact, the higher temperature scale indeed shows the $d$-wave pairing instability hosted by $d_{z^2}$ orbital. And the eigenvectors for the hole-doped ($n=2.8$) and electron-doped ($n=3.2$) cases are similar to those in the undoped case and are presented in the Appendix.

As shown in the upper panel of Fig.~\ref{value}, the eigenvalues $\lambda_{s^\pm}$ for the $s^\pm$-wave pairing at the three characteristic dopings $n = 2.8, 3.0,$ and $3.2$ are not significantly affected by the potential bias, although a decreasing trend can be discerned as the field increases at each doping. This weak field dependence can be understood as follows. Since the $s^\pm$-wave pairing in our limited high temperature region considered here is primarily governed by the $d_{z^2}$ orbital, which will be discussed in more detail later, and the orbital occupancy difference $\Delta n_z$ shown in Fig.~\ref{filling} (a) exhibits much smaller change than $\Delta n_x$, the interlayer pairing strength consequently becomes less sensitive to the potential bias, at least within our temperature range.

In contrast, the $d$-wave eigenvalues $\lambda_{d_x}$ for the $d_{x^2-y^2}$ orbital, presented in the lower panels of Fig.~\ref{value}, exhibit a clear increase with the potential bias and show a considerably stronger field dependence than the $s^\pm$-wave case. Moreover, although this sensitivity becomes stronger as $n$ increases from $2.8$ to $3.2$, the eigenvalues show a slower enhancing trend with lowering the temperature, which implies the possible lower $T_c$ for $d$-wave pairing at electron dopings, which contradicts with the $s^{\pm}$-wave case.  Besides, as shown in Fig.~\ref{filling} (b), the top layer's occupancy $n_x^t$ at the three dopings increases nearly linearly with the potential bias. Despite that the extrapolated $T_c$ cannot be reliably obtained due to the limited temperature range, we conjecture a dome-shaped dependence of $T_c$ on the top-layer $d_{x^2-y^2}$ orbital occupancy, reminiscent of the single-band Hubbard model~\cite{Maier2019}, in which $T_c$ increases from the underdoped to the optimally doped region and then decreases in the overdoped regime.

\begin{figure*}
\psfig{figure=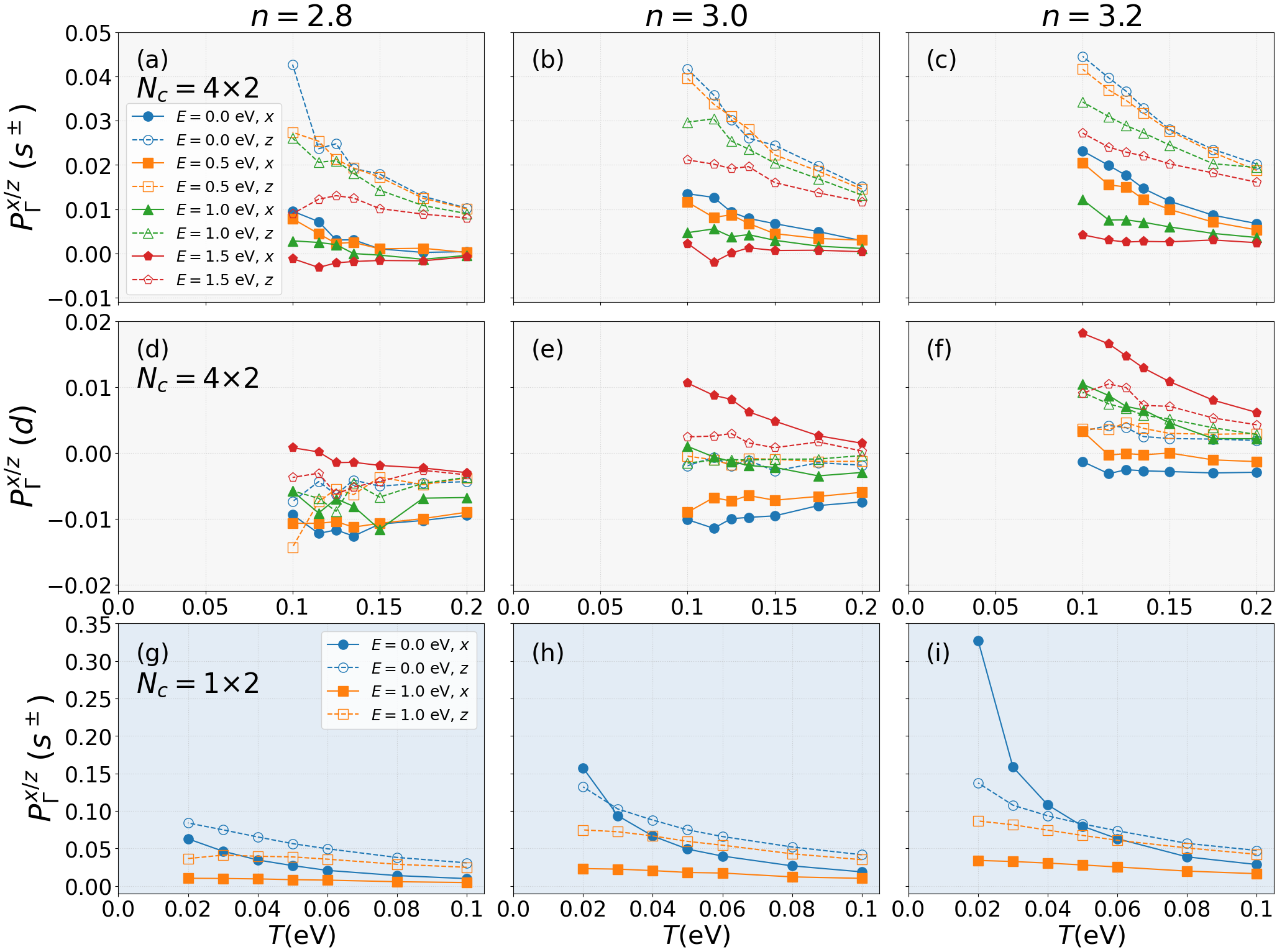,width=0.9\textwidth,clip}
\caption{
Temperature-dependent orbital-resolved interacting pair-field susceptibility $P^{x/z}_\Gamma(T)$ for $s^\pm$-wave (upper) and $d$-wave (middle) pairings, with $d_{x^2-y^2}$ (filled) and $d_{z^2}$ (hollow) orbitals, at potential bias $E = 0.0, 0.5, 1.0, 1.5$ eV for the $N_c = 4 \times 2$ cluster. The bottom row displays $P^{x/z}_\Gamma(T)$ for $s^\pm$-wave pairing of $N_c = 1 \times 2$ cluster, only including potential bias $E = 0.0, 1.0$ eV. Results are shown for hole-doped ($n = 2.8$), undoped ($n = 3.0$), and electron-doped ($n = 3.2$) cases (left to right).
}
\label{PGammaNc4}
\end{figure*}

\begin{figure*}
\psfig{figure=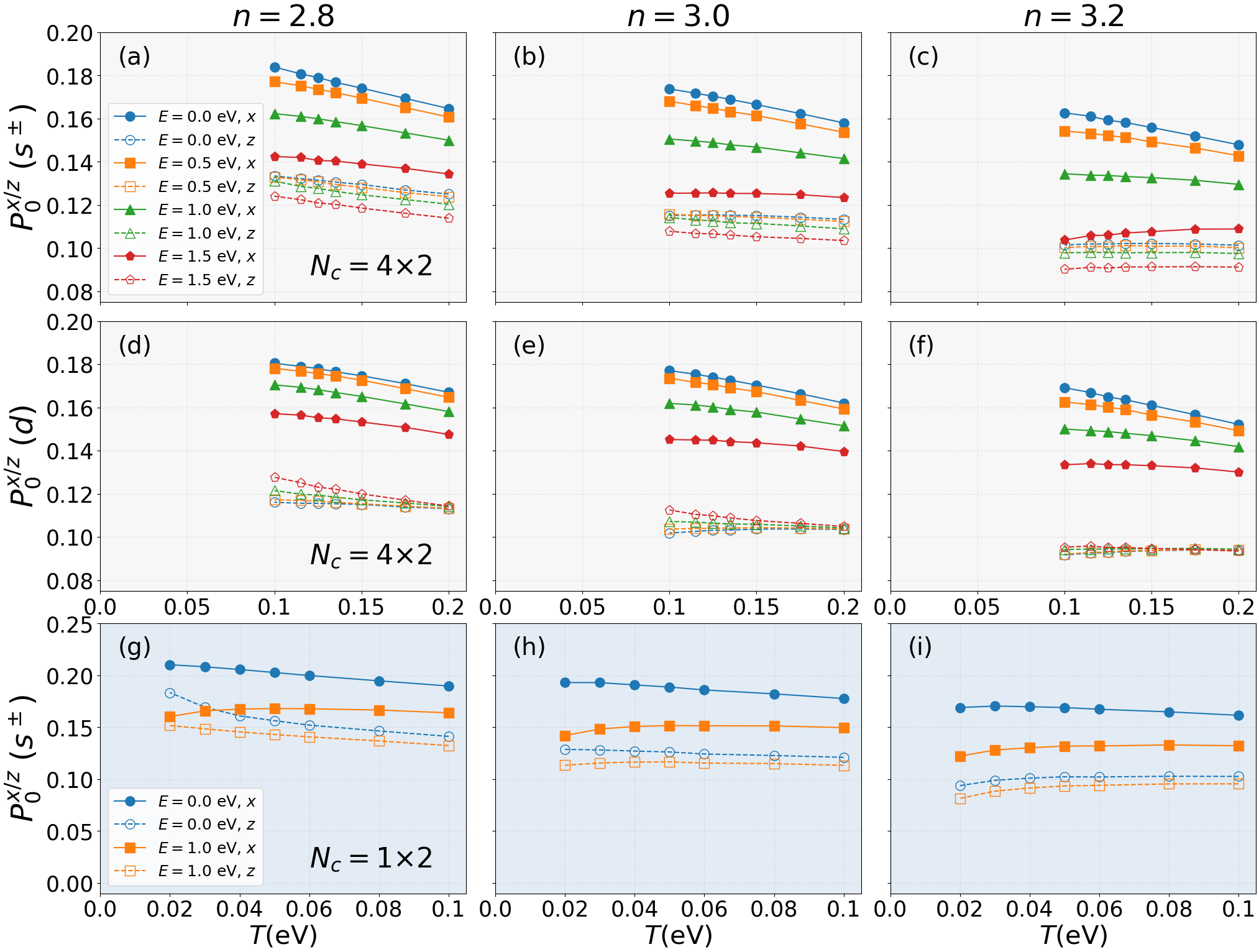,width=0.9\textwidth,clip}
\caption{
Akin to Fig.~\ref{PGammaNc4}, but for the temperature-dependent non-interacting pair-field susceptibility $P^{x/z}_0(T)$.}
\label{P0Nc4}
\end{figure*}

\subsection{Possible pairing transition via $P^{x/z}(T)$}
The orbital-resolved interacting and non-interacting pair-field susceptibility offers a more detailed understanding of a possible pairing transition induced by the layer potential bias. 
The upper panels of Fig.~\ref{PGammaNc4} presents the temperature evolution of the interacting $P_\Gamma^{x/z}(s^\pm)$ for $s^{\pm}$-wave. 
At each doping, increasing the potential bias $E$ suppresses both the $d_{z^2}$ and $d_{x^2-y^2}$ orbital contributions, with the $d_{z^2}$ component remaining consistently larger than its $d_{x^2-y^2}$ counterpart. This observation is in agreement with the BSE results, which show that the $d_{z^2}$ orbital dominates the $s^\pm$-wave pairing in the high temperature regime. 
Furthermore, in the absence of the potential bias, the pairing tendency is enhanced in the electron-doped case shown in Fig.~\ref{PGammaNc4} (c) compared to the hole-doped case shown in Fig.~\ref{PGammaNc4} (a), and this feature becomes more pronounced in $N_c=1\times2$ (Fig.~\ref{PGammaNc4} (g)(h)(i)) at lower temperatures. Therefore, we suggest that the $s^\pm$-wave pairing can be enhanced via electron doping, as has been proposed in previous works~\cite{YF2,YF2026,Edoping,WW2026}. 

More importantly, the $N_c = 1 \times 2$ cluster results reveal that the $d_{x^2-y^2}$ orbital dominates the $s^\pm$-wave pairing at lower temperatures. This finding suggests that the $d_{x^2-y^2}$ orbital component $P_\Gamma^{x}(s^\pm)$ (solid line in Fig.~\ref{PGammaNc4}) in the higher-temperature regime of the $N_c = 4 \times 2$ cluster deserves closer scrutiny. Besides, this $P_\Gamma^{x}(s^\pm)$ is suppressed by the potential bias for all three doping cases.

In the middle panels of Fig.~\ref{PGammaNc4}, the $d$-wave pairing on the $d_{x^2-y^2}$ orbital is apparently enhanced by the potential bias for all three dopings, with the strongest enhancement occurring in the electron-doped case. This behavior is consistent with the expectation that increased occupancy of the $d_{x^2-y^2}$ orbital towards the half-filling shown in Fig.~\ref{filling} could enhance antiferromagnetic fluctuations, which are known to favor $d$-wave pairing. However, the non-interacting contribution $P_0^{x/z}(d)$ in the middle panel of Fig.~\ref{P0Nc4} imply that the electric-field-induced increase of the $d_{x^2-y^2}$ orbital density suppresses its pairing mobility. We therefore propose that there exists a dome-shaped $T_c$ for the $d$-wave pairing, as discussed above for Fig.~\ref{value}. Meanwhile, the non-interacting $P_0^{x/z}(s^\pm)$ in the upper panel of Fig.~\ref{P0Nc4} behaves similar to $P_\Gamma^{x/z}(s^\pm)$, with both orbital components suppressed by the potential bias, indicating a steady suppression of the $s^\pm$-wave pairing with increasing field. The comparison between $E=0$ and $1.0$ eV for the $N_c = 1 \times 2$ cluster, presented in Fig.~\ref{P0Nc4} (g)(h)(i), further corroborates this conclusion at lower temperatures for all three doping cases.

As a result, with the $s^\pm$-wave pairing suppressed and the $d$-wave pairing enhanced, a possible electric-field-induced pairing-symmetry transition is plausible, consistent with previous SBMF results in the strong-coupling limit~\cite{YFNC}. 
In addition, the transition from $s^\pm$-wave to $d$-wave should be robust against the variation of $J$ within physical range. The main role of $J$ is to transfer the interlayer superexchange from $d_{z^2}$ to $d_{x^2-y^2}$ orbital to induce interlayer $s^\pm$-wave pairing in the latter~\cite{WCJ2024,YFarxiv,YFNC}. Therefore, stronger (weaker) $J$ is associate with stronger (weaker) interlayer $s^\pm$-wave pairing, and the electric field required to suppress such pairing is stronger (weaker). However, even under the strong-$J$ limit case as reported in Ref.~\cite{YFNC}, a realistic strength of the electric field is possible to induce the transition from $s^\pm$-wave to $d$-wave SC. Furthermore, the $d$-wave SC state under strong electric field is insensitive to the strength of $J$. Nonetheless, the critical field $E_c$ for this transition cannot be precisely determined in our calculations due to computational difficulty such as the fermionic sign problem. This underscores the need for more advanced numerical techniques capable of accessing lower temperatures to fully resolve the pairing transition.


\section{summary and outlook}
Motivated by the idea that a perpendicular electric field can modify the density distribution in the two-orbital bilayer model and thereby alter the superconducting pairing symmetry to potentially enhancing $T_c$, we employed the large-scale dynamical cluster quantum Monte Carlo calculations to study the bilayer two-orbital model in the undoped, hole-doped, and electron-doped cases with an additional layer potential bias. 

Our results show that the $s^\pm$-wave pairing, which is rooted in the $d_{z^2}$ orbital at high temperatures but may shift to the $d_{x^2-y^2}$ orbital at lower temperatures, is weakened by the density discrepancy induced by the additional potential bias. Furthermore, the potential bias induced $d$-wave pairing on the $d_{x^2-y^2}$ orbital may exhibit a dome-shaped dependence, reminiscent of the behavior observed in the single-band Hubbard model. Consequently, a pairing-symmetry transition driven by the potential bias is possible. Meanwhile, our analysis of the pair-field susceptibility indicates that the $s^\pm$-wave pairing symmetry can be enhanced by electron doping~\cite{YF2,Edoping,YF2026} without the potential bias, a finding that warrants further detailed investigation.

Although our method captures the essential features of an imbalanced bilayer two-orbital model mimicking the La$_3$Ni$_2$O$_7$ thin film applied with a perpendicular electric field, the accessible temperatures are limited by the fermionic sign problem, preventing us from reaching sufficiently low temperatures. Other possibilities include $s^\pm$-wave pairing hosted by the $d_{x^2-y^2}$ orbital~\cite{WCJ2024,YFarxiv}, which may lead to different doping-dependent behavior, such as a half-dome in $T_c$ under hole-doping and a saturated $T_c$ under electron-doping~\cite{halfdome}. Further investigation using more advanced many-body computational methods is desirable to clarify the enhanced $d$-wave pairing owing to the density distribution and other relevant possibilities.
Additionally, we remark that although our potential bias required for the $d$-wave SC may be considerably large for experiments on thin-film La$_3$Ni$_2$O$_7$, it does not lie entirely beyond the reach of current experimental capabilities~\cite{bigE1,bigE2}. Overall, we anticipate that future electric-field experiments on nickelates will provide valuable insights.

\setcounter{figure}{0}
\renewcommand{\thefigure}{A\arabic{figure}}
\begin{figure}
\psfig{figure=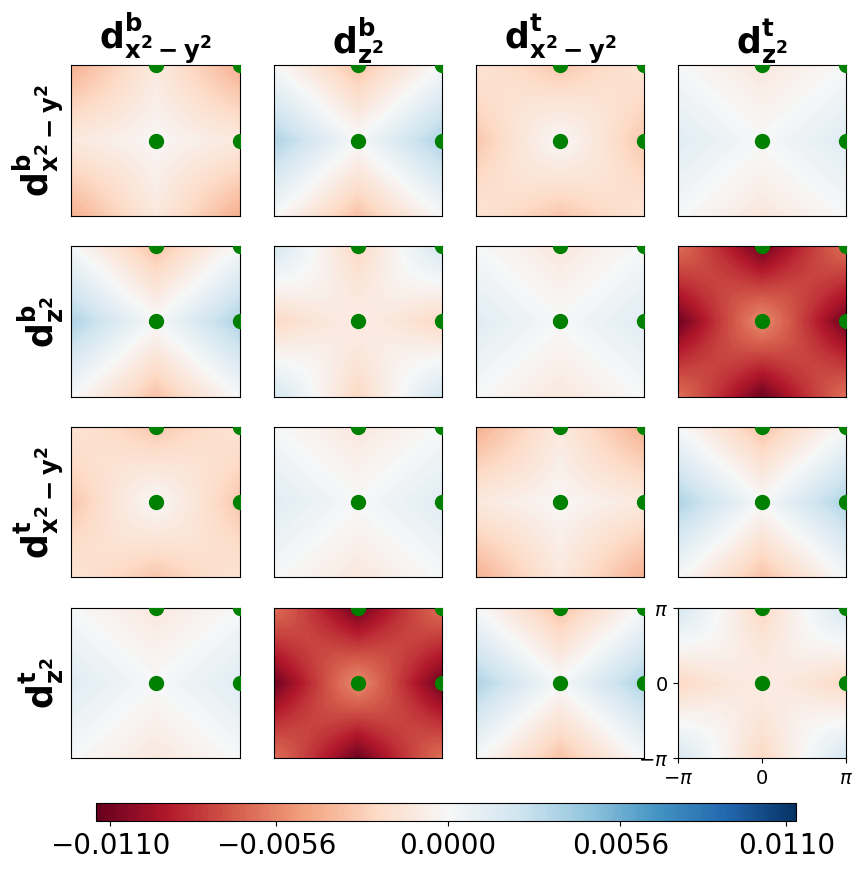,width=0.23\textwidth,clip}
\psfig{figure=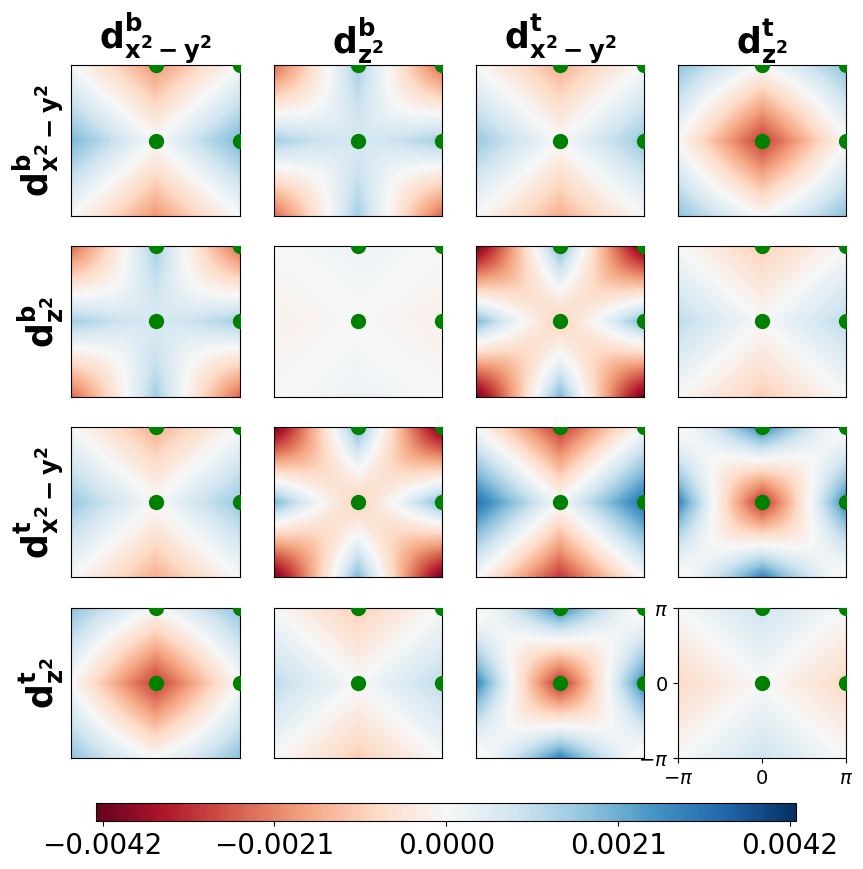,width=0.23\textwidth,clip}
\psfig{figure=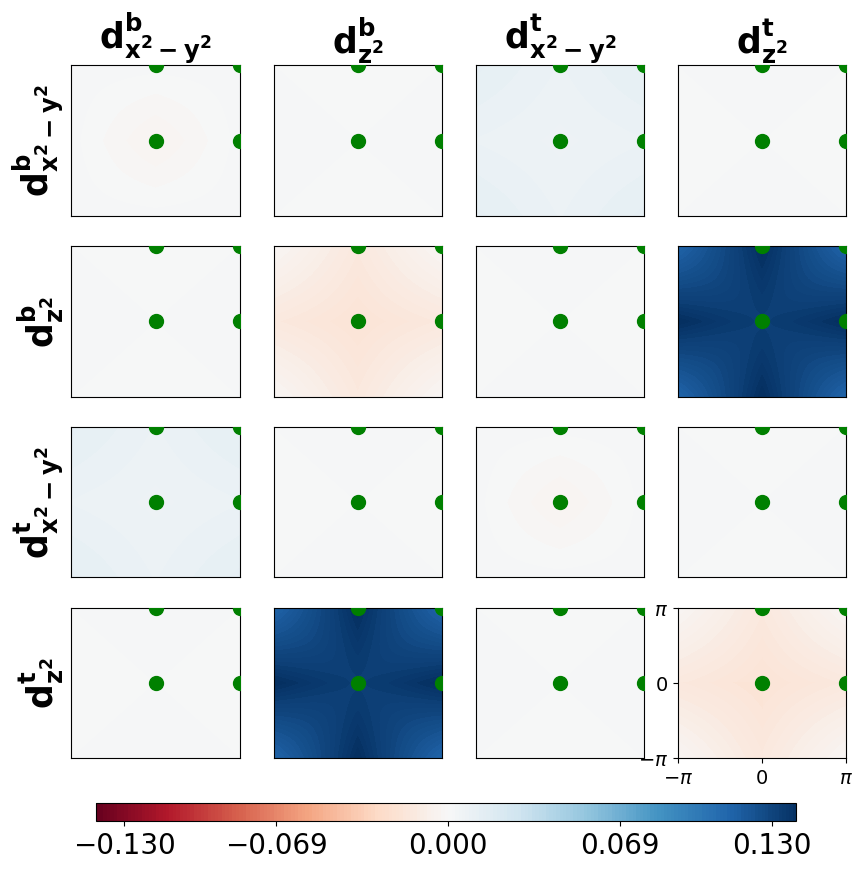,width=0.23\textwidth,clip}
\psfig{figure=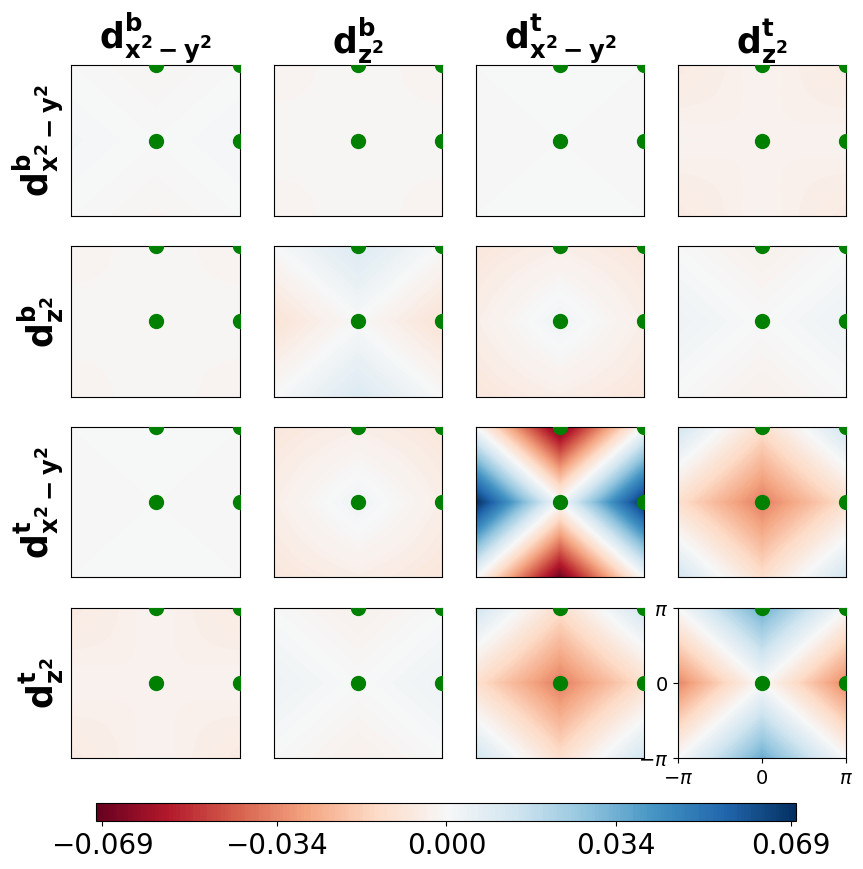,width=0.23\textwidth,clip}
\caption{The leading BSE eigenvector in orbital space at $E = 0.0$ (left) and $1.0$ eV (right) at $T = 0.1$ eV in the hole-doped ($n=2.8$, upper) and electron-doped ($n=3.2$, lower) cases, which are similar to the pairing symmetry transition from $s^\pm$-wave to $d$-wave in undoped $n=3.0$ case.}
\label{vectordoped}
\end{figure}

\section{Acknowledgment}
We acknowledge the support of the National Natural Science Foundation of China (Grant Nos.12174278, 12574141, and 12234016) and State Key Laboratory of Surface Physics and Department of Physics in Fudan University (Granandt No.KF2025\_12).
Y. Wei acknowledges the Undergraduate Training Program for Innovation and Entrepreneurship, Soochow University (Grant No.202510285040).
Mi Jiang also acknowledges the support of the China Scholarship Council (CSC) and the hospitality of Karsten Held at TU Wien.

\begin{figure}[H]
\psfig{figure=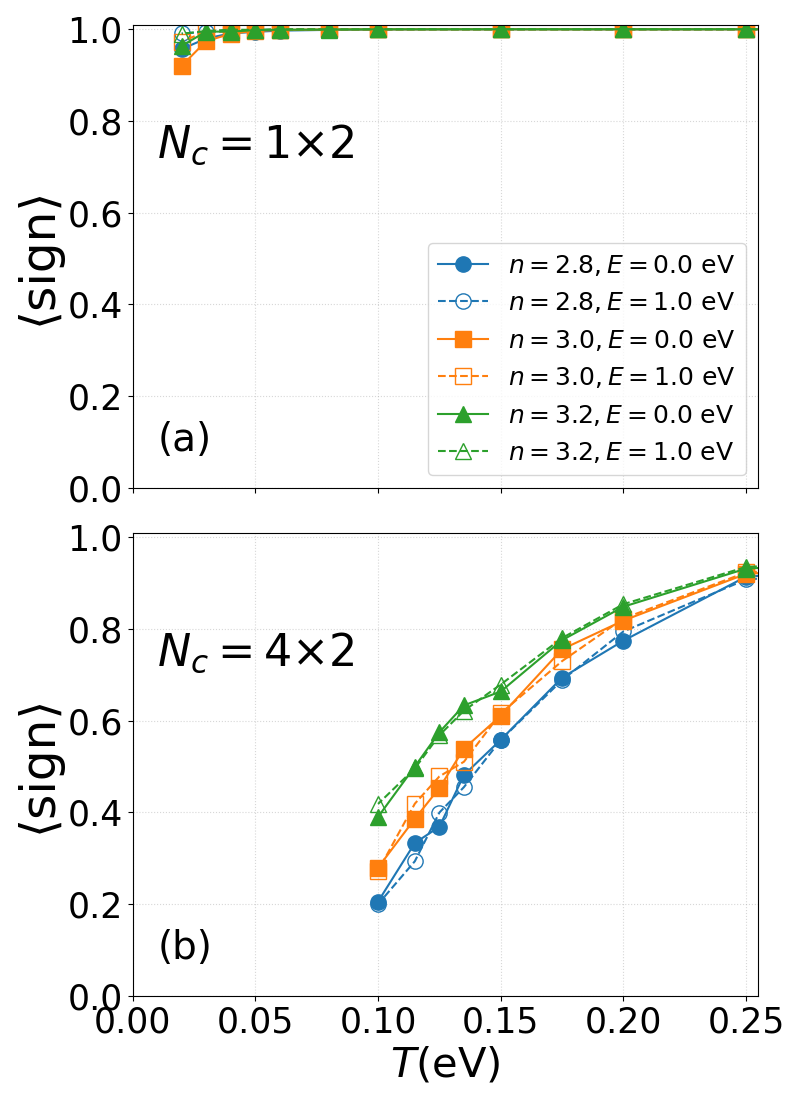,width=0.45\textwidth,clip}
\caption{
Temperature dependence of the fermionic sign problem on the $N_c=1\times2$ and $N_c=4\times2$ cluster for the three representative doping levels at $E=0.0$ and $1.0$ eV.}
\label{sign}
\end{figure}

\section{appendix}
\subsection{Eigenvectors for hole and electron doped cases}

We supplement the main text with the eigenvectors of doped systems. As shown in Fig.~\ref{vectordoped}, similar to the undoped case presented in Fig.~\ref{vectorn3} of the main text, $s^\pm$-wave pairing at $E=0$ and $d$-wave pairing at $E=1.0$ eV, is also seen in the hole-doped (upper) and electron-doped (lower) cases.

\subsection{Numerical details}

In our dynamical cluster approximation (DCA), we use the continuous-time auxiliary-field (CT-AUX) quantum Monte Carlo (QMC) method as the cluster solver~\cite{GullCTAUX}. The fermionic frequency grid contains 256 Matsubara frequencies for single-particle quantities and 32 frequencies for four-point correlation functions. For the Monte Carlo sampling, we perform 500 warm-up sweeps for thermalization, followed by $5\times10^6$ measurements with two sweeps per measurement. And the statistical accuracy is controlled by the total number of measurements rather than the number of walkers. These choices allow for efficient simulations within our available computational resources. 

As shown in Fig.~\ref{sign} (a), the fermionic sign problem is almost absent for the $N_c = 1 \times 2$ cluster. However, the sign problem becomes significantly more severe for the $N_c = 4 \times 2$ cluster, worsening with both decreasing temperature and decreasing electron density, as displayed in Fig.~\ref{sign} (b), indicating that larger cluster calculations are infeasible. In addition, the sign problem remains largely insensitive to the applied potential bias $E$, and we adopt a conservative threshold of $\langle \text{sign} \rangle \ge 0.2$ to ensure reliable measurements of observables.


\bibliography{main}

\end{document}